\documentclass[conference]{IEEEtran}
\IEEEoverridecommandlockouts
\usepackage{cite}
\usepackage{amsmath,amssymb,amsfonts}
\usepackage{algorithmic}
\usepackage{graphicx}
\usepackage{textcomp}
\usepackage{xcolor}
\def\BibTeX{{\rm B\kern-.05em{\sc i\kern-.025em b}\kern-.08em
    T\kern-.1667em\lower.7ex\hbox{E}\kern-.125emX}}
\usepackage{graphicx}
\usepackage{subcaption}
\graphicspath{ {images/} }
\usepackage[utf8]{inputenc}
\usepackage{fancyhdr}
\usepackage{dirtytalk}
\usepackage{tabularx}
\PassOptionsToPackage{hyphens}{url}\usepackage{hyperref}
\usepackage[ruled,lined,linesnumbered]{algorithm2e}

\usepackage{enumitem,kantlipsum}

\usepackage{graphicx}
\graphicspath{ {images/} }
\usepackage[utf8]{inputenc}
\usepackage{fancyhdr}
\usepackage{dirtytalk}
\usepackage{tikz}
\usetikzlibrary{shapes.geometric, arrows}
\usepackage{pgfplots}
\pgfplotsset{width=8cm,compat=1.16}
\usepgfplotslibrary{external}
\usepackage{mathtools}

\tikzstyle{startstop} = [rectangle, rounded corners, minimum width=2cm, minimum height=0.5cm,text centered, draw=black, fill=red!30]
\tikzstyle{process} = [rectangle, minimum width=2cm, minimum height=0.5cm, text centered, draw=black, fill=orange!30, align=left]
\tikzstyle{decision} = [diamond, minimum width=1.0cm, minimum height=0.4cm, text centered, draw=black, fill=green!30]
\tikzstyle{arrow} = [thick,->,>=stealth]

\newcolumntype{L}{>{$}l<{$}}

\usepackage[labelformat=simple]{subcaption}

\begin{document}


\title{Enhancing Trust and Security in the Vehicular Metaverse: A Reputation-Based Mechanism for Participants with Moral Hazard}


\author{\IEEEauthorblockN{Lotfi Ismail\IEEEauthorrefmark{1},
Marwa Qaraqe\IEEEauthorrefmark{1},
Ali Ghrayeb\IEEEauthorrefmark{2},
Dusit Niyato\IEEEauthorrefmark{3}}

\IEEEauthorblockA{\IEEEauthorrefmark{1} Division of Information and Computing Technology, College of Science and Engineering,\\
Hamad Bin Khalifa University, Qatar Foundation, Doha, Qatar.}
\IEEEauthorblockA{\IEEEauthorrefmark{2}  Electrical and Computer Engineering (ECE)
department, Texas A\&M University at Qatar, Doha, Qatar.}
\IEEEauthorblockA{\IEEEauthorrefmark{3}School of Computer Science and Engineering,
Nanyang Technological University, Singapore}
}

\maketitle
\begin{abstract}

In this paper, we tackle the issue of moral hazard within the realm of the vehicular Metaverse. 
A pivotal facilitator of the vehicular Metaverse is the effective orchestration of its market elements, primarily comprised of sensing internet of things (SIoT) devices. These SIoT devices play a critical role by furnishing the virtual service provider (VSP) with real-time sensing data, allowing for the faithful replication of the physical environment within the virtual realm.
However, SIoT devices with intentional misbehavior can identify a loophole in the system post-payment and proceeds to deliver falsified content, which cause the whole vehicular Metaverse to collapse.
To combat this significant problem, we propose an incentive mechanism centered around a reputation-based strategy. 
Specifically, the concept involves maintaining reputation scores for participants based on their interactions with the VSP. These scores are derived from feedback received by the VSP from Metaverse users regarding the content delivered by the VSP and are managed using a subjective logic model.
Nevertheless, to prevent ``good" SIoT devices with false positive ratings to leave the Metaverse market, we build a vanishing-like system of previous ratings so that the VSP can make informed decisions based on the most recent and accurate data available. 
Finally, we validate our proposed model through extensive simulations. Our primary results show that our mechanism can efficiently prevent malicious devices from starting their poisoning attacks. At the same time, trustworthy SIoT devices that had a previous miss-classification are not banned from the market.

\end{abstract}


\section{Introduction}


The Metaverse, regarded as the next generation of the internet, is spreading recently over several applications such as remote workplaces, vehicular systems and online banking, amongst others~\cite{Minrui_COMST_2023, ISMAIL_JSAC_2023_semantic}. Several technologies such as virtual reality (VR), augmented reality (AR) and artificial intelligence (AI) are key enablers of the Metaverse. With the release of Meta's Quest 3, mixed reality (XR), another key enabler of the Metaverse, is closer to realization than ever before\footnote{https://www.meta.com/quest/quest-3/}.
As the majority of the Metaverse users are expected to be mobile, e.g., vehicular Metaverse, next-generation mobile networks should be well designed to support the different Metaverse services.

The first step towards exploring the benefits of vehicular Metaverse is the rendering of the physical world into its digital twin. For this objective, the VSP recruits sensing IoT (SIoT) devices (e.g., UAVs, intelligent cars and CCTV cameras) in the field to collect data about the physical environment and then deliver that data to the VSP. Nevertheless, as the majority of this collected data consists of images and video scenes, the network load increases exponentially and reaches its limit. Additionally, appropriate mechanisms should be designed to incentivize SIoT devices to share their data with the VSP.
To address these challenges, existing work proposed the use of semantic communication to reduce the data load and enable the VSP to collect more data with high freshness levels~\cite{Ismail_FNWF_2022, ISMAIL_JSAC_2023_semantic, Jiacheng_TWC_2022}. The idea of semantic communication consists of applying machine learning (ML) algorithms to extract the semantic information from the raw data then delivers only the extracted information to the VSP. To incentivize SIoT devices, auction theory and contract theory were adapted to design mechanisms that  guarantee the properties of individual rationality (IR) and incentive compatibility (IC)~\cite{Ismail_FNWF_2022, ISMAIL_JSAC_2023_semantic}.

However, the integrity of the delivered data might be unsecured due to the moral hazard problem. Although incentive mechanisms with IR and IC properties guaranteed prevent adverse selection problem, the moral hazard problem remains unsolved. While the adverse selection problem appears before the agreement on the contract between the two parties (i.e., the VSP and the SIoT devices), the moral hazard problem appears after the agreement. Specifically, the SIoT device can deliver content that does not meet the requirements of the VSP as agreed initially. With the emergence of adversarial attacks~\cite{Zhang_CCS_2023}, it becomes difficult to differentiate between legitimate content and malicious content. Importantly, in the scenario of vehicular Metaverse, a malicious SIoT device can launch a poisoning attack on the delivered semantic data to cause perturbation of the Metaverse system, which can degrade the quality of service (QoS) and the quality of experience (QoE) of the Metaverse users significantly. Even more concerning is the potential for life-threatening attacks capable of triggering accidents in the physical world~\cite{Minrui_COMST_2023}.

Although the Metaverse has attracted the communication and networking researchers recently, a number of works are already available to address a number of security issues in these systems.
In~\cite{Zhang_JSAC_2023}, a moving target defense (MTD) strategy was proposed to secure digital twins over mobile networks where the network characteristics are altered to disrupt various phases of the cyber kill chain. 
The problem is then formulated using a semi-Markov decision process and solved using a deep reinforcement learning algorithm.
The idea behind MTD schemes consists of dynamically modifying relevant network properties, such as transmission routes and IP addresses, to disrupt the adversary's prior knowledge. However, the problem addressed in~\cite{Zhang_JSAC_2023} is quite different from our focus in this work. Specifically, the main problem addressed in~\cite{Zhang_JSAC_2023} is related to preventing distributed denial-of-service (DDoS) attacks while our focus is on mitigating poisoning attacks on vehicular Metaverse from SIoT devices with moral hazard.

To this end, the problem of moral hazard in the vehicular Metaverse should be urgently addressed before their deployment. To the best of our knowledge, no previous work has addressed the problem of moral hazard in Metaverse services. We address this gap by proposing a greedy algorithm based on subjective logic and reputation mechanism to incentivize SIoT devices to participate in the Metaverse services truthfully at all times.
In summary, our main contributions are as follows:
\begin{itemize}
    \item  We introduce a novel approach to address the issue of moral hazard within the context of the vehicular Metaverse, providing a unique perspective on trust and security concerns in this emerging technology. The proposed reputation-based mechanism effectively deters dishonest behavior by participants, enhancing the overall integrity of the system.

    \item We use subjective logic to model the reputation of each SIoT device and introduce a vanishing-like system to manage previous ratings. This innovative approach helps prevent reputable SIoT devices from leaving the market due to occasional false positive ratings.

    \item The extensive simulations conducted to validate the proposed model provide empirical evidence of its effectiveness. This empirical validation reinforces the practical applicability and robustness of the proposed mechanism.
\end{itemize}













\section{System Model}
\subsection{System Model Description}
\begin{figure}[ht!]
    \centering
    \includegraphics[width=.45\textwidth,height=4.5cm]{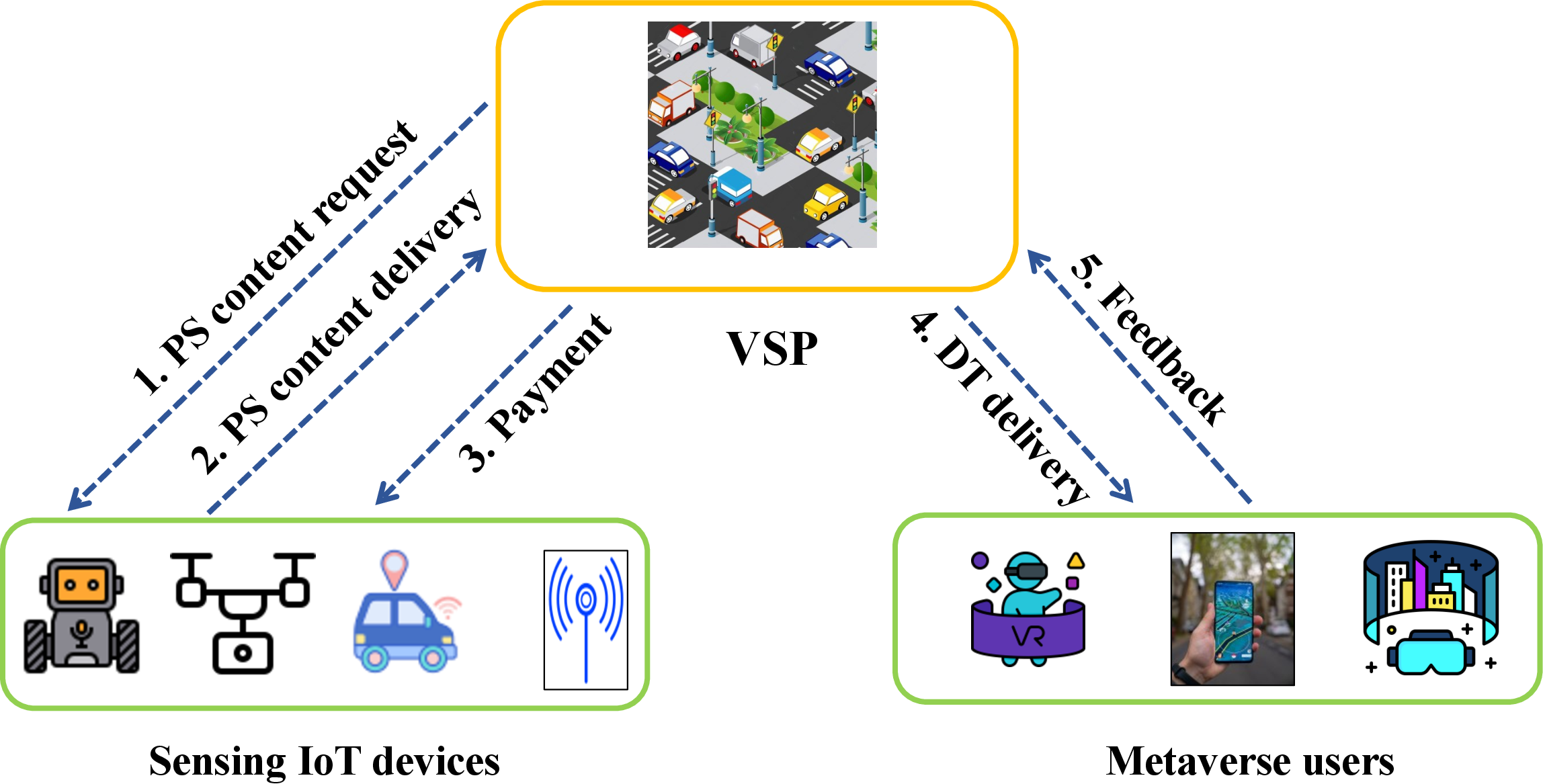}
    \caption{System model of anti-poisoning attacks vehicular Metaverse ecosystem.}
    \label{fig:conf_system_model}
\end{figure}

Fig.~\ref{fig:conf_system_model} describes our proposed system model. 
We examine a Metaverse market comprising a VSP, a set of $\mathcal{N} = \{1, \dots, i, \dots, N\}$ IoT devices, acting as data owners and a set of $\mathcal{M} = \{1, \dots, j, \dots, M\}$ vehicular Metaverse users (VMUs).
The VSP's role is to construct the digital twin within the Metaverse using data gathered from the set of chosen SIoT devices and then deliver the construct digital twin to the VMUs. Each SIoT device is equipped with a suite of sensors for gathering geo-spatial data from its immediate surroundings, along with machine learning (ML) models designed to extract semantic information from the collected raw data. The SIoT devices send the extracted semantic data to the VSP instead of all the raw data which brings more efficiency to the network load~\cite{Ismail_FNWF_2022}. 
Given multiple sellers (representing the IoT devices) and a single buyer (the VSP), the choice of a reverse auction, which essentially reverses the traditional roles of buyers and sellers, aligning well with our system design~\cite{Nisan_2007}.
Nevertheless, an auction mechanism that guarantees an optimal solution and satisfies both IR and IC conditions only addresses the problem of adverse selection which prevents a malicious participant from selecting a bundle not dedicated for them. The auction mechanism still faces another major security threat where participants with moral hazards can inject a poisoning attack into the Metaverse system. The VSP should be able to prevent such attacks in real-time to avoid degradation of its services.

In what follows, we first start by highlighting the major type of poisoning attacks that can be launched by malicious SIoT devices with moral hazard. We then present an overview of our proposed system design to build a robust Metaverse system that can efficiently reduce the occurrence of such poisoning attacks.

\subsection{Poisoning Attack Models}
The attackers are considered to follow a randomized strategy to attack the vehicular Metaverse system. Specifically, a malicious SIoT device alternates between sending legitimate semantic data and sending falsified semantic data to lure the VSP.
Different types of poisoning attacks can be launched by malicious SIoT devices. Here, we highlight three major types of attacks:
\begin{itemize}
    \item \textbf{Reducing service cost:} As the continuous collection of sensing data and the execution of semantic data extraction algorithms require continuous energy consumption, a selfish SIoT device can choose to deliver semantic data that is based on outdated sensing data. This action will cause the malicious SIoT device to increase its utility (as lower energy consumption is required, see~\eqref{eq:u_i}) while the rendered digital twin will not reflect real-time dynamics of the physical world, causing degraded QoE of the VMUs.

    \item \textbf{Digital twin tampering:} While the previous type of attack does not have an objective of causing damages to the Metaverse services, the digital twin tampering is initiated by a malicious SIoT device to cause severe degradation of the Metaverse services. For instance, after extracting the semantic from real-time images of the physical world, the attacker alters the position or the type of certain object so as the experience of the Metaverse users is degraded to the maximum. Interestingly, if the actions taken in the digital twin are reflected in the physical world, serious safety issues can arise, e.g., car accidents.
    Fig.~\ref{fig:conf_semantci_attack} illustrates an instance of such attack.

    \begin{figure}[ht!]
        \centering
        \includegraphics[width=.4\textwidth,height=4.0cm]{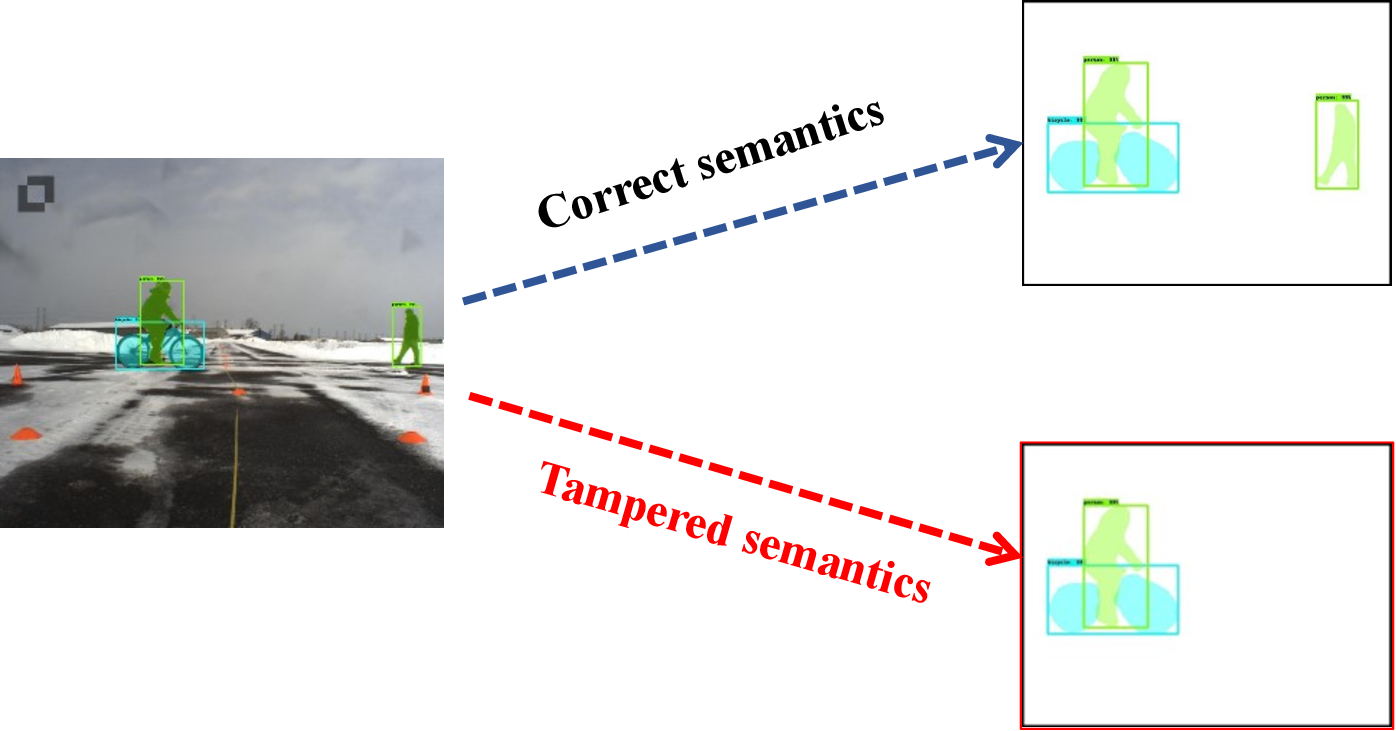}
        \caption{Exemplary case of semantic attack where the attacker hides the person on the right side of the image.}
        \label{fig:conf_semantci_attack}
    \end{figure}

    \item \textbf{Man-in-the-middle attack:} This attack can be launched from an SIoT device with low security defense levels, e.g., data encryption is compromised. The attacker transmits their altered data instead of the legitimate SIoT device data. The VSP has no visibility of the existence of the attacker in the middle of the transmission. Therefore, if the attacker is detected, the compromised SIoT device is flagged as an untrustworthy device.
\end{itemize}

\subsection{System Overflow}
\begin{itemize}
    \item First, as described in Fig.~\ref{fig:conf_system_model}, the VSP initiates a request to collect physical world data from SIoT devices in the field using a reverse auction mechanism. Interested SIoT devices located in the region of the VSP' interest express their willingness to sell their data with their offered prices. Specifically, SIoT device $i$ reports its type $t_i =\{b_i, \hat{s}_i\}$ to the VSP, where $b_i$ indicates the price offered for its data and $\hat{s}_i$ represent the semantic value of the offered data. The semantic value of the data, $\hat{s}_i$, is calculated as in~\cite{Ismail_FNWF_2022} and includes parameters such as weather condition, number of detected objects and their positions. 
    For instance, in adverse weather conditions, the valuation of the derived semantic information is expected to rise, indicating its heightened significance to the VSP.
    Additionally, as SIoT devices require a specific number of channels to deliver their semantic data based on their size, each SIoT device $i$ also reports $C_i$, the number of channels required, to the VSP. Later, we show how the VSP uses this information when selecting the set of winners (i.e., the set of SIoT devices who won the auction and are allowed to deliver their data to the VSP). 

    \item After receiving the offers from all SIoT devices, the VSP first uses the reputation database it has constructed over the previous interactions with SIoT devices and VMUs to decide which SIoT devices are more likely to be trustworthy, which are denoted by $\mathcal{N}^{'}$. The set of $\mathcal{N}^{'}$ SIoT devices is then used in the winner selection algorithm to decide the final list of SIoT devices from which the VSP buys the semantic data with the given payments.

    \item Once the VSP receives the semantic data from the chosen SIoT devices, the digital twin is then constructed in the vehicular Metaverse and then delivered to the VMUs. Based on the quality of service experienced by the VMUs, feedback ratings are sent by the VMUs to the VSP to update the reputation score of all participating SIoT devices. In the next round of semantic data collection, the VSP uses the updated reputation scores to select trustworthy SIoT devices. 
\end{itemize}


\section{Problem Formulation And Methods}


In this section, we first present the formulation of reputation management for SIoT devices utilizing subjective logic. Subsequently, we detail our approach to effectively filter reputations obtained from the VMUs. Finally, we formulate the reverse auction mechanism as a social welfare maximization problem.

\subsection{Subjective Logic Model for Reputation Calculation}

We use $r_{i:j}=\{b_{i:j}, d_{i:j}, u_{i:j}, \alpha_{i}\}$ to denote reputation score received from Metaverse user $j$ to SIoT device $i$.
$b_{i:j}$, $d_{i:j}$, $u_{i:j}$ and $\alpha_{i}$ represent belief, disbelief, uncertainty, and the effective degree of uncertainty effect on reputation of SIoT device $i$, respectively~\cite{Kang_IoTJ_2019_Reputation, Oren_subjctv_2007}. 
$b_{i:j}+ d_{i:j}+ u_{i:j} = 1$ and $b_{i:j}$, $d_{i:j}$, $u_{i:j}$, $\alpha_{i} \in [0,1]$.
We define $R_{i:j}$ as the reputation of SIoT device $i$ given by VMU $j$, which is given by
\begin{equation}
    R_{i:j} = b_{i:j} +\alpha_{i}u_{i:j}.
\end{equation}

As the VSP interacts with SIoT devices and get feedback from VMUs over several time periods, the local reputation for an SIoT device $i$ constructed at the VSP is expressed based on this period. We denote $\tau$ as the effective period of interaction and divide it into $Y$ time periods, i.e., $\{t_1, \dots, t_y, \dots, t_Y\}$. The reputation value of SIoT device $i$ based on feedback from VMU $j$ at timestep $t_y$ is then calculated as

{\small \begin{equation}\label{eq:reputation_1}
\begin{cases} 
b^{t_y}_{i:j} = \frac{\omega_1 p^{t_y}_{i:j}}{\omega_1 p^{t_y}_{i:j} + \omega_2 q^{t_y}_{i:j}+\kappa},
\\ d^{t_y}_{i:j} = \frac{\omega_2 q^{t_y}_{i:j}}{\omega_1 p^{t_y}_{i:j} + \omega_2 q^{t_y}_{i:j}+\kappa},
\\ u^{t_y}_{i:j} = \frac{\kappa}{\omega_1 p^{t_y}_{i:j} + \omega_2 q^{t_y}_{i:j}+\kappa}.
\end{cases}
\end{equation}}

\noindent where $p^{t_y}_{i:j}$ and $q^{t_y}_{i:j}$ represent the number of positive and negative interactions between the SIoT device $i$ and VMU $j$, respectively. $\omega_1$ and $\omega_2$ are weighting factors for positive and negative interactions, respectively, and are considered to sum-up to 1, i.e., $\omega_1+\omega_2=1$. $\kappa$ is a constant value that reflects the rate of uncertainty and is set to $\kappa=1$~\cite{Kang_IoTJ_2019_Reputation, Oren_subjctv_2007}.

Motivated by the fact that recent interactions where malicious devices attempt to attack the Metaverse system are more important than older interactions, we consider the following formulation to show their importance in calculating the final reputation score. Recent interaction events hold greater significance due to their freshness, carrying a higher weight than those from the past. Specifically, we define the fading function that reflects this behavior as $ \psi(t_y) = \psi_y = z^{Y-y}$, where $z\in(0,1)$ is a given fading parameter and $y$ is the current time slot that determines the freshness degree of the reputation value. Therefore, \eqref{eq:reputation_1} is reformulated as

{\small \begin{equation}\label{eq:reputation_2}
\begin{cases} 
b^{fin}_{i:j} = \frac{\sum^{Y}_{y=1}\psi_y b^{t_y}_{i:j}}{\sum^{Y}_{y=1}\psi_y},
\\ d^{fin}_{i:j} = \frac{\sum^{Y}_{y=1}\psi_y d^{t_y}_{i:j}}{\sum^{Y}_{y=1}\psi_y},
\\ u^{fin}_{i:j} = \frac{\sum^{Y}_{y=1}\psi_y u^{t_y}_{i:j}}{\sum^{Y}_{y=1}\psi_y}.
\end{cases}
\end{equation}}

\noindent Therefore, the final reputation value of SIoT device $i$ from VMU $j$ is calculated as $R^{fin}_{i:j} = b^{fin}_{i:j} +\alpha_{i}u^{fin}_{i:j}$.
Finally, we calculate the final reputation value of a SIoT device $i$ as the average value of the reputation values received from all VMUs, i.e., 
\begin{equation}\label{eq:R_fin}
    R^{fin}_{i} = \frac{\sum^M_{j=1}R^{fin}_{i:j}}{M}.
\end{equation}

\subsection{Reputation Backpropagation}
\begin{figure}[ht!]
    \centering
    \includegraphics[width=.45\textwidth,height=4.5cm]{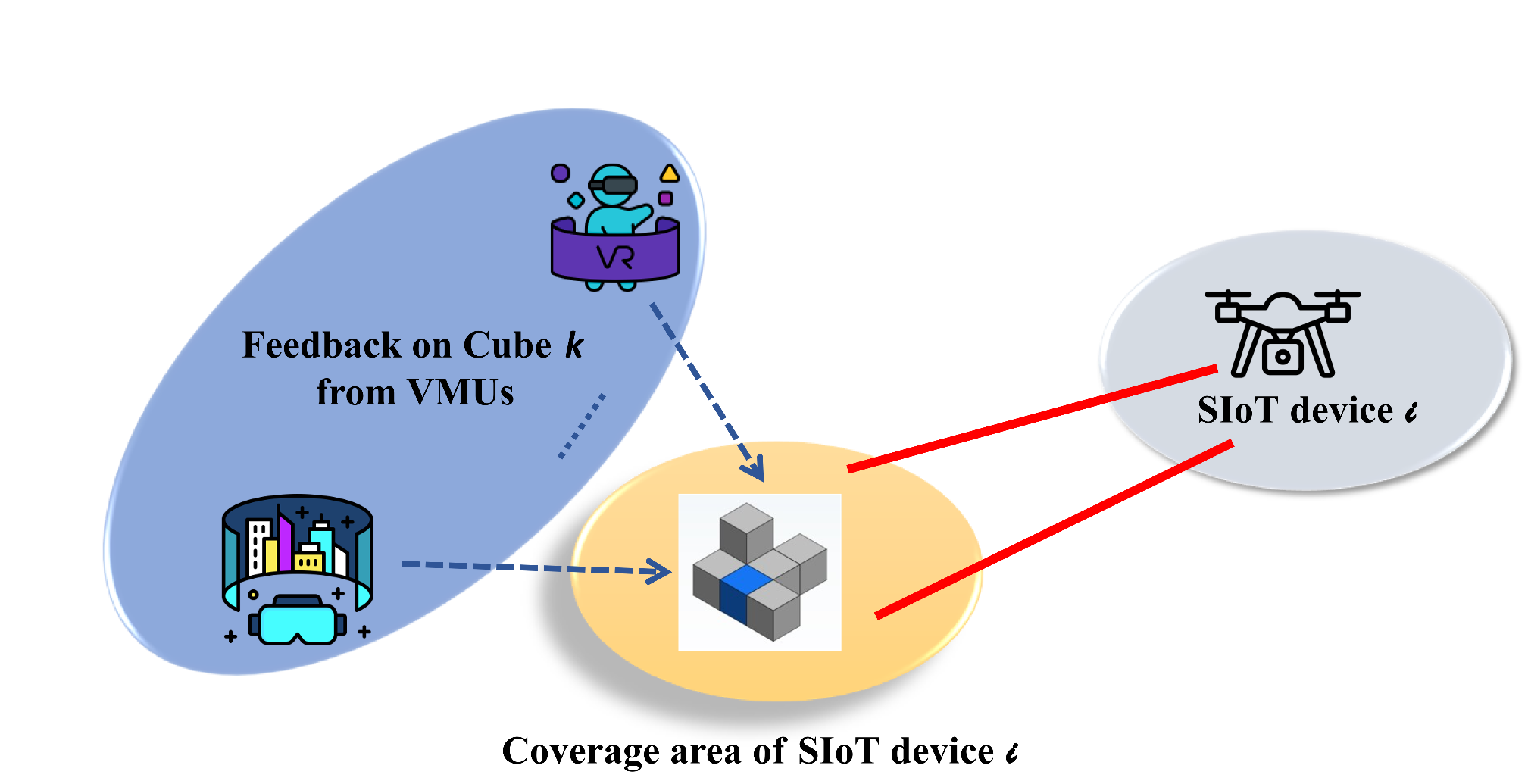}
    \caption{Representation of the cube under interest.}
    \label{fig:conf_backpropagation}
\end{figure}

One of our key contributions in this paper is the \emph{reputation backpropagation mechanism} which is based on the human-in-the-loop design and is presented in Fig.~\ref{fig:conf_backpropagation}. Specifically, as there is no direct mapping between the VMUs and the SIoT devices, an important issue to solve is how to backpropagate the feedback received by a VMU at the VSP server to a specific SIoT device. Motivated by the idea of backpropagation in deep neural networks (DNN)\cite{Goodfellow_2016}, we develop a new approach to backpropagate the feedbacks received by the VSP from the VMUs. The reputation backpropagation is conducted as follows:

\begin{itemize}
    \item \textbf{Step 1 } \textit{3D Grid :} As the area of coverage by the $\mathcal{M}$ SIoT devices is regarded as a 3D grid containing $K$ cubes, VMUs send feedbacks on regions of the grid in which they interacted with. We denote the subset of SIoT devices who covered cube $k$ as $\mathcal{S}_k \subset \mathcal{M}$.

    \item \textbf{Step 2 } \textit{Negative feedbacks only:} However, as the space of interaction between the VMUs and the Metaverse interface is large (e.g., touch screen), we consider that the VMUs send negative feedbacks only on regions, i.e., cubes in the 3D grid, in which they experienced a low QoE. Therefore, not receiving a feedback about the other cubes is considered as a positive feedback in the reputation mechanism. 


    \item \textbf{Step 3 } \textit{Feedback Filtering :} However, a low QoE in a specific cube is likely attributed to the misbehavior of just one of the SIoT devices, rather than all of them. Consequently, some SIoT devices may receive an inaccurate negative rating. To mitigate the impact of this issue, reputations undergo a final filtering process outlined in Algorithm~\ref{algo:feedback_filtering}.
\end{itemize}

\begin{algorithm}[ht!]
\SetAlgoLined
\SetKwInOut{Input}{Input}
\SetKwInOut{Output}{Output}
 \Input{Initialize an empty set $\mathcal{N}^{-}$ for SIoT devices with negative rating;   }
 \Output{The filtered set of SIoT devices, $\mathcal{N}^{-}$, which deserves a negative rating;}
 \SetKwBlock{Beginn}{beginn}{ende}
    \Begin{

    \ForEach{$k \in K$}{ 
        Calculate pairwise similarity of the content delivered by SIoT devices in the subset $\mathcal{S}_k$;\\
        SIoT device with the lowest similarity score with the others is added to $\mathcal{N}^{-}$;\\
    }
    SIoT devices \textbf{not in }$\mathcal{N}^{-}$ are assigned a positive reputation;\\
    
    
    }
 \caption{Feedback Filtering Algorithm}
 \label{algo:feedback_filtering}
\end{algorithm}





\subsection{Social Welfare Maximization Reverse Auction}

The social welfare, defined as the system efficiency~\cite{Zhang_2017_welfare}, is a key metric in evaluating the effectiveness of the Metaverse market system (i.e., selling the semantic data by the SIoT devices to the VSP). Therefore, maximizing social welfare signifies an efficient operation of the system.
The major objective of auction-based mechanism design is to prevent market manipulation by malicious SIoT devices which can lead to an unjustified increase in their utilities.
Detailed description of our developed reverse auction mechanism that addresses these issues are omitted here and can be found in~\cite{Ismail_FNWF_2022}.
Here, we briefly define the utility of the SIoT devices and the VSP to formulate the social welfare of the auction mechanism, which is necessary for the performance evaluation.

The SIoT device $i$ utility is defined as
\begin{equation}\label{eq:u_i}
    u_i = p_i - c_i,
\end{equation}
\noindent where $p_i$ is the payment received from the VSP, and $c_i$ is the service cost which covers the communication cost, data collection cost and execution of the semantic extraction algorithms.
The VSP's utility is defined as 
\begin{equation}\label{eq:u_hat}
    \hat{u} = \sum\limits_{i \in \mathcal{N}}\xi_i R_i^S  - \sum\limits_{i\in \mathcal{N}}\xi_i p_i - \hat{c},
\end{equation}

\noindent where $\xi_i$ is a binary variable denoting whether a SIoT device $i$ is selected amongst the winners or not, $R_i^S$ is the value of the semantic data delivered to the VSP by SIoT device $i$ and $\hat{c}$ is the cost for allocation the wireless channel from spectrum service providers by the VSP. The social welfare of the system is quantified as the aggregate of utilities derived from all entities within the system, including both the VSP and the SIoT devices and is written as
\begin{equation}
    S(\xi) = \sum\limits_{i\in \mathcal{N}}\xi_i \left( R_i^S - c_i \right) - \hat{c}.
\end{equation}    

Finally, the social welfare maximization problem can be written as an integer linear programming (ILP) subject to the availability of channels as follows
\begin{subequations}
\label{eq:optz_1}
\begin{align}
\begin{split}
\max_{\xi} S(\xi) = \sum\limits_{i\in \mathcal{N}}\xi_i \left( R_i^S - c_i \right) - \hat{c}, \label{eq:MaxA} 
\end{split}\\
\begin{split}
\hspace{1cm} s.t. \sum_{i\in \mathcal{N}} \xi_i C_i \leq B \label{eq:MaxB}
\end{split}\\
\begin{split}
\hspace{1cm} \xi_i \in \{0,1\}, \forall i\in \mathcal{N} \label{eq:MaxC}
\end{split}
\end{align}
\end{subequations}

\noindent where $C_i$ is the number of channels requested by SIoT device $i$ and $B$ represents the number of channels offered by the VSP. The ILP presented in~\eqref{eq:optz_1} can be straightforwardly solved using a deterministic off-the-shelf solver, e.g., \emph{Gurobi optimizer}.

\section{Experiments}
In this section, we perform experiments using real-world data and provide numerical results to assess the effectiveness of our proposed reputation-based incentive mechanism for poisoning attack mitigation.
Unless otherwise stated, we consider that the number of SIoT devices is $N=20$ and that the channel's capacity for each SIoT device is fixed to $r= 10 kbps$.

\subsection{Dataset and Simulations Settings}
We carry out our experiments using the CARRADA dataset~\cite{Ouaknine_CARRADA_2020}, a contemporary open dataset featuring 30 scenes of synchronized sequences of camera and radar images. The radar images encompass both the range-angle (RA) and the range-Doppler (RD) views.
More description on the dataset can be found in~\cite{Ismail_FNWF_2022} and are omitted here for brevity.
The output of the semantic segmentation algorithms yields an image with a white background, featuring the masked objects (refer to Fig.~\ref{fig:conf_semantci_attack}). Additionally, a meta-data text file in JSON format is generated. This file encompasses a mapping of every object within the image, along with its associated class, as well as additional derived semantic attributes like range and shape.
The meta-data text file has a maximum size of $d_{meta} = 1 kb$.
To reflect the the real environment where different SIoT devices capture the same scene but from different angles, we consider that SIoT devices transmit a set of scenes that are very close in time (in $ms$). At each episode, a different scene from the $30$ synchronized sequences is used\footnote{The implementation of the 3D grid representation is left for the future work.}.
In addition, we consider that the simulations last for $10$ episodes where each episode has a length of $50$ time-steps. To model the different types of SIoT devices that participate in the vehicular Metaverse market, we group the SIoT devices into three types:
\begin{itemize}
    \item \textbf{Type-1 SIoT:} at every episode, an attack is launched with probability $p=0.05$ in each timestep. This type reflects normal SIoT device with no willingness to attack and no man-in-the-middle attack.

    \item \textbf{Type-2 SIoT:} at every episode, an attack is launched with probability $p=0.5$ in each timestep.

    \item \textbf{Type-3 SIoT:} at the beginning of every episode, choose with probability $p=0.5$ either to attack for the whole episode or not to attack.
\end{itemize}

\noindent Finally, the types of the participating SIoT devices are evenly distributed.

\subsection{Results}

\begin{figure}[ht!]
    \centering
    \includegraphics[width=.45\textwidth,height=4.5cm]{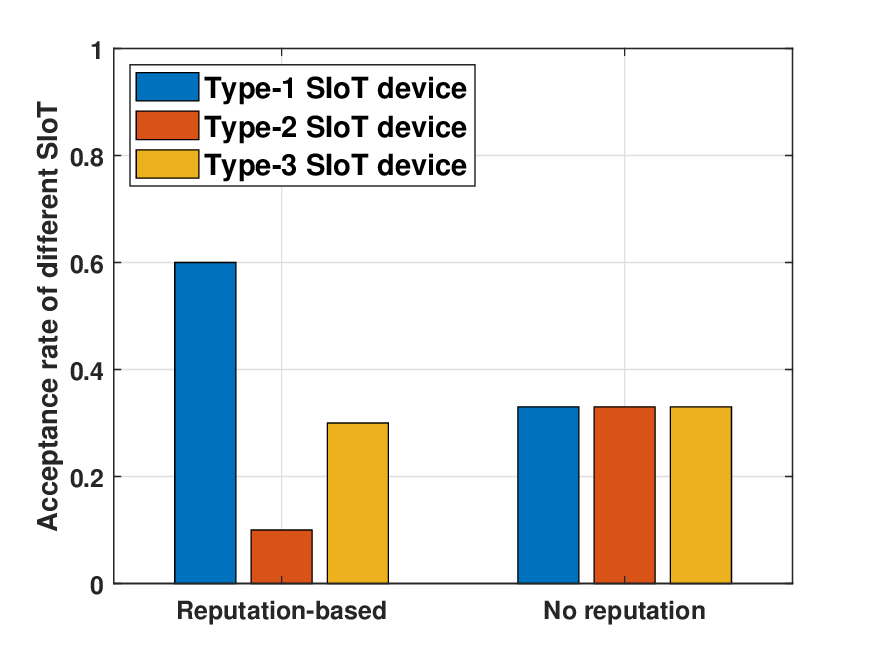}
    \caption{Acceptance rate of SIoT devices from different types with and without reputation mechanism.}
    \label{fig:conf_result_1}
\end{figure}

We first start by comparing the case when no anti-poisoning attack is used and the case where our algorithm is used, which is illustrated in Fig.~\ref{fig:conf_result_1}. We observe that with the use of our technique, the frequency of tampered digital twin decreases significantly. In the case without reputation mechanism, the VSP chooses SIoT devices from the different types with equal numbers. However, when our developed reputation-based anti-poisoning attack mechanism is used, SIoT devices dominate the set of selected SIoT devices for semantic data delivery. Interestingly, we observe that the SIoT devices from \textbf{type-3} have a higher percentage compared to SIoT devices from \textbf{type-2}. We explain this behavior by the fact that SIoT devices from \textbf{type-3} choose to be non-malicious for certain episodes for the whole period. By observing this behavior in real-time, the VSP is able to benefit from these SIoT devices and collect their semantic data. This further shows the benefits of collecting real-time feedbacks from VMUs.

\begin{figure}[ht!]
    \centering
    \includegraphics[width=.45\textwidth,height=4.5cm]{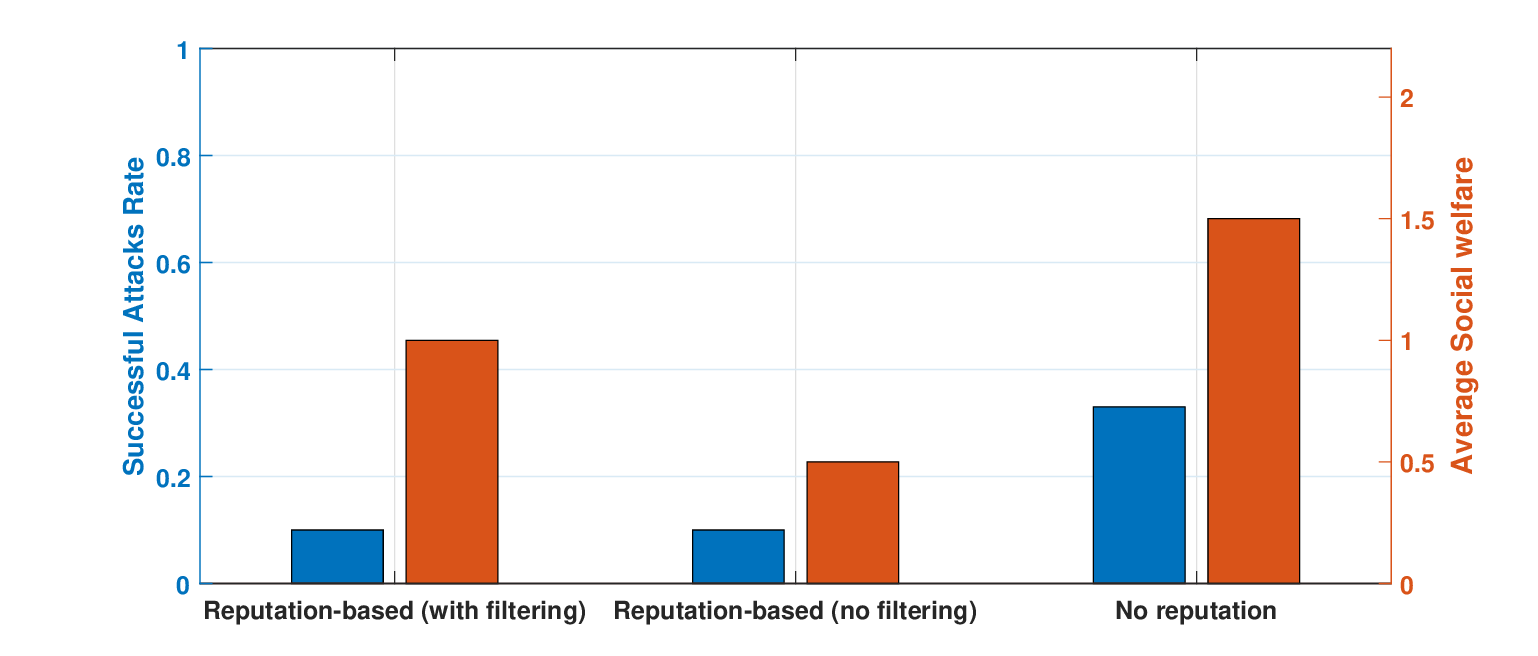}
    \caption{Average social welfare and successful attack rates in different scenarios.}
    \label{fig:conf_result_2}
\end{figure}

Next, to observe the impact of the reputation backpropagation filtering Algorithm on our anti-poisoning mechanism, we plot in Fig.~\ref{fig:conf_result_2} social welfare of the system averaged over all the episodes and the successful attack rate for three cases: the case where no anti-poisoning attack mechanism is adopted (denoted by ``No reputation"), the case where our anti-poisoning attack mechanism is used with and without the reputation backpropagation is adopted. We observe that the average social welfare of the system without the reputation mechanism is much higher than that when our reputation-based mechanism is adopted, which seems to be an undesirable result. Higher value of the social welfare is expected to reflect a higher system efficiency. This behavior is justified by the fact that our anti-poisoning mechanism does not address the social welfare maximization issue. Specifically, we should first note that the social welfare captures only the system benefits with respect to the defined utility functions in~\eqref{eq:u_i} and~\eqref{eq:u_hat} without considering decrease in the security performance (i.e., digital twin tampering). This is clearly captured by the high successful attack rate in the case where no anti-poisoning attack mechanism is used. As shown in Fig.~\ref{fig:conf_result_1}, our proposed anti-poisoning attack mechanism decreases the successful attack rate on the system significantly.

Finally, we also observe from Fig.~\ref{fig:conf_result_2} the advantages of our feedback filtering algorithm on the performance of our system. Specifically, although the successful attack rate in the case where the filtering is used and that where it is not used is the same, we observe a significant decrease in the average social welfare of the system. This argument is supported by the notion that when negative feedback is distributed regarding specific regions in the delivered Metaverse content, non-malicious SIoT devices (SIoT devices with false negative ratings) may be deterred from allocation in subsequent iterations for content delivery. This is reflected in the decrease of the value of the semantic data as formulated in~\eqref{eq:optz_1}.





\section{Conclusion And Future Works}


In conclusion, this paper addresses the critical issue of moral hazard within the domain of the vehicular Metaverse. By proposing an incentive mechanism rooted in a reputation-based strategy, we aim to deter dishonest actors from engaging in fraudulent behavior after receiving payments. This approach involves assigning reputation scores to participants based on their interactions with the VSP, informed by feedback from Metaverse users.
Furthermore, we implement a vanishing-like system for previous ratings to prevent reputable SIoT devices with occasional false positive ratings from exiting the market. This ensures that the VSP can make decisions based on the most timely and accurate data available, upholding the integrity of the Metaverse market and enabling trustworthy IoT devices to continue contributing their sensing data.
Ultimately, extensive simulations validate the effectiveness of our proposed model. The results demonstrate the efficiency of our mechanism in preventing malicious devices from launching poisoning attacks, while also ensuring that reliable SIoT devices with prior missclassifications are not unfairly excluded from the market. This research provides a robust framework for enhancing security and trustworthiness in the vehicular Metaverse ecosystem.
For future works, it is worth exploring the direction of identifying the ideal weights for diminishing the influence of previous data.


\bibliographystyle{IEEEtran}
\bibliography{reference}

\end{document}